# Generation of polarization-entangled Bell states in monolithic photonic waveguides by leveraging intrinsic crystal properties

**Trevor G. Vrckovnik**[1,2,3], **Dennis Arslan**[1], **Falk Eilenberger**[1,2,3], and **Sebastian W. Schmitt**[1,2,*]

[1]Fraunhofer Institute for Applied Optics and Precision Engineering IOF, Albert-Einstein-Str. 7, 07745 Jena, Germany
[2]Institute of Applied Physics, Abbe Center of Photonics, Friedrich Schiller University Jena, Albert-Einstein-Str. 15, 07745 Jena, Germany
[3]Max Planck School of Photonics, Albert-Einstein-Str. 15, 07745 Jena, Germany
[*]E-mail: sebastian.wolfgang.schmitt@iof.fraunhofer.de

**Advanced photonic quantum technologies—from quantum key distribution to quantum computing—require on-chip sources of entangled photons that are both efficient and readily scalable. In this theoretical study, we demonstrate the generation of polarization-entangled Bell states in structurally simple waveguides by exploiting the intrinsic properties of nonlinear crystals. We thereby circumvent elaborate phase-matching strategies that commonly involve the spatial modulation of a waveguide's linear or nonlinear optical properties. We derive general criteria for the second-order susceptibility tensor that enable the generation of cross-polarized photon pairs via spontaneous parametric down-conversion in single-material waveguides. Based on these criteria, we systematically categorize all birefringent, non-centrosymmetric crystal classes in terms of their suitability. Using coupled mode theory, we then numerically analyze cuboid waveguides made from two materials that are highly relevant to integrated photonics: lithium niobate, a well-established platform, and barium titanate, an emerging alternative. We find that barium titanate consistently outperforms lithium niobate by providing a higher nonlinear efficiency and high concurrence over a significantly broader spectral range. These findings outline a practical route toward highly efficient, fabrication-friendly, and scalable sources of polarization-entangled photons for integrated quantum photonic circuits.**

## 1 Introduction

The rapid advancements in quantum information processing and optical telecommunications have significantly increased the demand for integrated nonlinear quantum photonic devices[1,2]. Many emerging quantum technologies—such as quantum key distribution (QKD) systems[3,4], quantum random number generators[5,6], quantum repeaters[7,8], and photonic quantum processors[9,10]—rely on entangled photon pairs as a fundamental resource. To enable compact, scalable, and stable implementations, it is essential to generate these photons directly on-chip, rather than using bulk crystals or optical fibers[1,2]. Depending on the application, different types of entanglement are required. Polarization entanglement is essential for quantum communication protocols like QKD[11], where qubits are encoded in polarization states. Time-bin entanglement offers robustness against decoherence in fiber networks and long-distance transmission[12]. Path entanglement plays a central role in integrated quantum computing by enabling interference-based operations[13], while frequency entanglement is increasingly used for multiplexed quantum communication[14]. Each form imposes distinct requirements on source design and integration, reinforcing the need for versatile and scalable on-chip entanglement generation.

The most widely used physical mechanism for the on-chip generation of entangled photons is spontaneous parametric down conversion (SPDC), the quantum analogue of sum frequency generation (SFG) and second harmonic generation (SHG)[15,16]. For these nonlinear processes to be efficient in waveguides, it is essential to utilize a material with a second-order susceptibility tensor $\chi^{(2)}$ that has large components at specific indices, and to achieve phase-matching between the pump, signal, and idler eigenmodes of the waveguide. The latter can be accomplished with a careful design of the waveguide geometry and by utilizing the dispersion and/or birefringence of a material's linear susceptibility $\chi^{(1)}$[17]. SPDC has been successfully utilized in integrated waveguides to generate frequency-entangled photon pairs, most notably mediated by quasi-phase matching in domain engineered lithium niobate[18,19] and modal phase matching in dispersion engineered semiconductor waveguides[20].

The generation of polarization-entangled photon pairs, on the other hand, remains far more challenging. To date, polarization entanglement has only been demonstrated in elaborate structures such as Bragg-reflection waveguides[21–24], periodically poled systems with spatial walk-off compensation[25,26], or hybrid integrated platforms combining multiple optical components and polarization rotators[27]. These approaches, while effective, introduce fabrication complexity and hinder scalability.

We address these technological problems by leveraging the intrinsic optical properties of nonlinear crystals in conjunction with a minimalist waveguide design: a crystalline cuboid core embedded in silicon dioxide ($SiO_2$).

More precisely, we first establish general requirements on the $\chi^{(2)}$ tensor for the generation of cross-polarized signal and idler modes in monolithic waveguides via SPDC, and then categorize all birefringent, non-centrosymmetric crystal classes according to their principal suitability. With this, we identify two promising material platforms for integrated photonic devices: lithium niobate[28,29] ($LiNbO_3$) and barium titanate[30,31] ($BaTiO_3$). Using our own numeric implementation of the coupled mode theory (CMT), capable of handling arbitrary $\chi^{(2)}$, we perform SHG simulations for various waveguide widths and heights to find the most efficient geometries, where we observe that $BaTiO_3$ achieves higher nonlinear efficiencies than $LiNbO_3$. Finally, SFG simulations of the most promising geometries allow us to infer the likelihood of generated photon pairs and their concurrence in the SPDC process via the quantum-classical correspondence principle, where $BaTiO_3$ exhibits a high concurrence over a larger bandwidth than $LiNbO_3$.

## 2 Entangled Photon States via SPDC

SPDC can be thought of as the time-reversed process of SFG, where a single input photon of a higher frequency (known as the pump), decomposes into two photons of lower frequencies (known as the signal and idler). Energy conservation of the process requires that

$$\omega_p = \omega_s + \omega_i \quad (1)$$

where $\omega_p$, $\omega_s$, $\omega_i$ are the angular frequencies of the pump, signal, and idler photon. As well as this inherent relationship of the frequencies of the signal-idler pair, there can also be relationships in their polarization, or for waveguides, the eigenmode they propagate in. For interactions occurring over long distances (such as in waveguides), conservation of momentum is also required

$$\omega_p\, n_{\text{eff}}^p = \omega_s\, n_{\text{eff}}^s + \omega_i\, n_{\text{eff}}^i \quad (2)$$

where $n_{\text{eff}}^p$, $n_{\text{eff}}^s$, and $n_{\text{eff}}^i$ are the effective refractive indices of the pump, signal, and idler eigenmodes. If we assume that the signal and idler photons can either be predominantly $x$-polarized (propagating in the fundamental TE mode) or predominantly $y$-polarized (propagating in the fundamental TM mode), then the two-qubit output state of the SPDC process can be expressed as

$$|\psi\rangle = a_{xx}|xx\rangle + a_{xy}|xy\rangle + a_{yx}|yx\rangle + a_{yy}|yy\rangle \quad (3)$$

where the first term in the ket vector represents the signal photon polarization, the second term the idler photon polarization, and the coefficients $a_{ij}$ are normalized to:

$$\langle\psi|\psi\rangle = |a_{xx}|^2 + |a_{xy}|^2 + |a_{yx}|^2 + |a_{yy}|^2 = 1 \quad (4)$$

Each of the four basis states in $|\psi\rangle$ represents a different possible emission channel for the SPDC process, and entanglement occurs when the specific channel used cannot be determined without measurement.

There are four states with maximum entanglement between the signal and idler polarization, known as the Bell states

$$|\Phi^\pm\rangle = \tfrac{1}{\sqrt{2}}\left(|xx\rangle \pm |yy\rangle\right) \quad (5)$$

$$|\Psi^\pm\rangle = \tfrac{1}{\sqrt{2}}\left(|xy\rangle \pm |yx\rangle\right) \quad (6)$$

where the $|\Phi^\pm\rangle$ Bell states can be created from $|\psi\rangle$ when the coefficients are set to $a_{xx} = \pm a_{yy}$ and $a_{xy} = a_{yx} = 0$. Similarly, the $|\Psi^\pm\rangle$ states can be created when $a_{xy} = \pm a_{yx}$ and $a_{xx} = a_{yy} = 0$.

## 3 Phase Matching Considerations

In waveguides, momentum conservation (Eq. (2)) is synonymous with phase matching, as the effective refractive indices determine both the photon momentum and the propagating phase of the eigenmodes. Since the effective refractive indices usually increase with frequency, it can be expected that phase matching in SPDC must occur between a higher-order pump mode and lower-order signal and idler modes. However, since higher-order modes are increasingly more sensitive to fabrication imperfections and harder to excite in a controlled manner, the order of all modes should be as low as possible to maximize mode overlap and avoid technical difficulties. The effective refractive indices can be specifically influenced by the waveguide geometry and the linear susceptibility $\chi^{(1)}$ of a material, where the orientation of a birefringent material, as opposed to an isotropic material, represents an additional degree of freedom in fine-tuning between modes of different polarization. Equations (1) and (2) can also be reinterpreted as a weighted arithmetic mean

$$n_{\text{eff}}^p = \frac{\omega_s\, n_{\text{eff}}^s + \omega_i\, n_{\text{eff}}^i}{\omega_s + \omega_i} \quad (7)$$

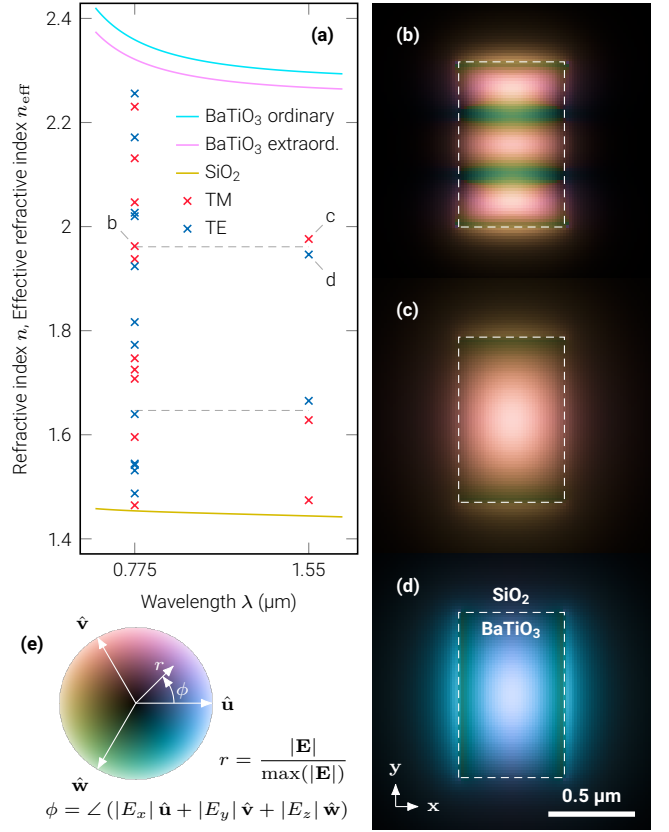

**Fig. 1 | The role of birefringence. (a)** The refractive index of $BaTiO_3$ along the ordinary and extraordinary axis, of $SiO_2$, and the effective refractive indices of predominantly $x$-polarized (TE) or $y$-polarized (TM) modes. Dashed horizontal lines illustrate the phase matching condition for the creation of $|\Psi^\pm\rangle$ Bell states. **(b–d)** The electric field profiles of the eigenmodes indicated in (a), where **(e)** is the associated color map. The waveguide is made of a $BaTiO_3$ core oriented such that the extraordinary axis $\tilde{z}$ is parallel to the $x$-axis of the waveguide. The core, outlined by dashed lines, has a width of $w = 0.61\,\mu m$ and a height of $h = 0.96\,\mu m$. The surrounding material is $SiO_2$.

where the angular frequencies represent the weights and are usually fixed, and the effective refractive indices must be adjusted to satisfy the equation.

If the signal and idler photons are to be generated in any of the four basis states of $|\psi\rangle$ (Eq. (3)), we can apply Eq. (7) to find the required $n_{\text{eff}}^p$ of the pump mode that achieves phase matching. For the generation of Bell states, two different basis states need to be simultaneously phase-matched. The required $n_{\text{eff}}^p$ for the $|xy\rangle$ and $|yx\rangle$ basis states are identical for $\omega_s = \omega_i$ (rather similar for $\omega_s \approx \omega_i$), which means that the $|\Psi^\pm\rangle$ Bell states can be phase-matched (rather well) with a single pump mode. However, the $|xx\rangle$ and $|yy\rangle$ basis states will generally require significantly different values of $n_{\text{eff}}^p$, and therefore no single pump mode can be phase-matched to the $|\Phi^\pm\rangle$ Bell states. The exception would be if the signal and idler modes are degenerate such that $n_{\text{eff}}^s = n_{\text{eff}}^i$. However, in this case the phase matching condition requires the same $n_{\text{eff}}^p$ for all four basis states, and therefore the $|\Phi^\pm\rangle$ Bell states can only emerge if $\chi^{(2)}$ forbids conversion into some of the basis states. Due to these severe fundamental limitations, we will focus in the remainder of this study on the efficient generation of $|\Psi^\pm\rangle$ Bell states.

Figure 1a illustrates how the birefringence and aspect ratio of a $BaTiO_3$ waveguide with a rectangular cross-section can be exploited to phase-match two fundamental modes of orthogonal polarization ($\lambda = 1.55\,\mu m$; cf. Figs. 1c and 1d) with a higher-order mode ($\lambda = 0.775\,\mu m$; cf. Fig. 1b). This means that the phase matching condition for the $|\Psi^\pm\rangle$ Bell states can be fulfilled for a single pump mode ($\lambda = 0.775$ µm), but not for the $|\Phi^\pm\rangle$ Bell states.

## 4 Polarization Entanglement in Waveguides

The polarization $\mathbf{P}$ of an anisotropic and nonlinear material, whose nonlinear response is approximately independent of the angular frequency $\omega$ of the exciting electric field $\mathbf{E}$, can be represented as[17]

$$\mathbf{P}(\omega) = \sum_{n=1}^{\infty} \mathbf{P}^{(n)}(\omega) \tag{8}$$

$$P_i^{(1)}(\omega) = \varepsilon_0 \sum_j \chi_{ij}^{(1)}(\omega)\, E_j(\omega) \tag{9}$$

$$P_i^{(2)}(\omega) = \varepsilon_0 \sum_{jk} \chi_{ijk}^{(2)} [E_j * E_k](\omega) \tag{10}$$

where $\varepsilon_0$ is the vacuum permittivity, $\boldsymbol{\chi}^{(n)}$ the $n$-th order susceptibility, and the convolution is defined as:

$$[f * g](\omega) = \int_{\mathbb{R}} f(\omega')\, g(\omega - \omega')\, \frac{\mathrm{d}\omega'}{2\pi} \tag{11}$$

We define a waveguide reference system $(x, y, z)$, where the positive $z$-axis represents the propagation direction of the waveguide and the positive $y$-axis points away from the waveguide substrate (compare Fig. 2n).

Since low-order modes tend to be mainly transversally polarized ($E_z(\omega) \approx 0$), it suffices to focus on the following components of Eq. (10)

$$\begin{pmatrix} P_x^{(2)}(\omega) \\ P_y^{(2)}(\omega) \end{pmatrix} = \varepsilon_0 \underline{\boldsymbol{\chi}}^{(2)} \cdot \begin{pmatrix} [E_x * E_x](\omega) \\ [E_y * E_y](\omega) \\ 2[E_x * E_y](\omega) \end{pmatrix} \tag{12}$$

where we shall call

$$\underline{\boldsymbol{\chi}}^{(2)} := \begin{pmatrix} \chi_{xxx}^{(2)} & \chi_{xyy}^{(2)} & \chi_{xxy}^{(2)} \\ \chi_{yxx}^{(2)} & \chi_{yyy}^{(2)} & \chi_{yxy}^{(2)} \end{pmatrix} \tag{13}$$

the reduced second-order susceptibility in the waveguide reference system. In the case where the electric field mainly consists of three harmonic waves

$$\mathbf{E}(t) = \mathrm{Re}\left(\mathbf{E}^{\mathrm{p}} e^{-\mathrm{i}\omega_{\mathrm{p}} t} + \mathbf{E}^{\mathrm{s}} e^{-\mathrm{i}\omega_{\mathrm{s}} t} + \mathbf{E}^{\mathrm{i}} e^{-\mathrm{i}\omega_{\mathrm{i}} t}\right) \tag{14}$$

where $\mathbf{E}^{\mathrm{p}}$, $\mathbf{E}^{\mathrm{s}}$, $\mathbf{E}^{\mathrm{i}}$ are the eigenmode profiles and $\omega_{\mathrm{p}}$, $\omega_{\mathrm{s}}$, $\omega_{\mathrm{i}}$ the angular frequencies of the pump, signal, and idler modes, the convolution terms evaluated at $\omega = \omega_{\mathrm{i}} = \omega_{\mathrm{p}} - \omega_{\mathrm{s}}$ can be written as

$$[E_m * E_n](\omega_{\mathrm{i}}) = \frac{1}{4}\left(E_m^{\mathrm{p}} \overline{E_n^{\mathrm{s}}} + E_n^{\mathrm{p}} \overline{E_m^{\mathrm{s}}}\right) \delta(\omega_{\mathrm{i}} - \omega_{\mathrm{p}} + \omega_{\mathrm{s}}) \tag{15}$$

where the overline denotes complex conjugation and $\delta(\cdot)$ is the Dirac delta. Then, by dropping the Dirac delta in favor of readability, Eq. (12) becomes:

$$\begin{pmatrix} P_x^{(2)}(\omega_{\mathrm{i}}) \\ P_y^{(2)}(\omega_{\mathrm{i}}) \end{pmatrix} = \frac{\varepsilon_0}{2} \underline{\boldsymbol{\chi}}^{(2)} \cdot \begin{pmatrix} E_x^{\mathrm{p}} \overline{E_x^{\mathrm{s}}} \\ E_y^{\mathrm{p}} \overline{E_y^{\mathrm{s}}} \\ E_x^{\mathrm{p}} \overline{E_y^{\mathrm{s}}} + E_y^{\mathrm{p}} \overline{E_x^{\mathrm{s}}} \end{pmatrix} \tag{16}$$

If the pump mode is predominantly $x$-polarized ($E_y^{\mathrm{p}} \approx 0$):

$$\begin{pmatrix} P_x^{(2)}(\omega_{\mathrm{i}}) \\ P_y^{(2)}(\omega_{\mathrm{i}}) \end{pmatrix} = \frac{\varepsilon_0}{2} \begin{pmatrix} \chi_{xxx}^{(2)} & \chi_{xxy}^{(2)} \\ \chi_{yxx}^{(2)} & \chi_{yxy}^{(2)} \end{pmatrix} \cdot \begin{pmatrix} \overline{E_x^{\mathrm{s}}} \\ \overline{E_y^{\mathrm{s}}} \end{pmatrix} E_x^{\mathrm{p}} \tag{17}$$

If the pump mode is predominantly $y$-polarized ($E_x^{\mathrm{p}} \approx 0$):

$$\begin{pmatrix} P_x^{(2)}(\omega_{\mathrm{i}}) \\ P_y^{(2)}(\omega_{\mathrm{i}}) \end{pmatrix} = \frac{\varepsilon_0}{2} \begin{pmatrix} \chi_{xxy}^{(2)} & \chi_{xyy}^{(2)} \\ \chi_{yxy}^{(2)} & \chi_{yyy}^{(2)} \end{pmatrix} \cdot \begin{pmatrix} \overline{E_x^{\mathrm{s}}} \\ \overline{E_y^{\mathrm{s}}} \end{pmatrix} E_y^{\mathrm{p}} \tag{18}$$

If the signal mode represents the quantum mechanical state of vacuum fluctuation, meaning the values of $E_x^{\mathrm{s}}$ and $E_y^{\mathrm{s}}$ are subject to randomness, but one still wants to enforce that the idler photon always arises in a polarization state orthogonal to the signal photon, which means that the output state is one of the $|\Psi^\pm\rangle$ Bell states, then the entries of $\underline{\boldsymbol{\chi}}^{(2)}$ must have specific values

$$\underline{\boldsymbol{\chi}}^{(2)} = \begin{cases} \begin{pmatrix} 0 & ? & \uparrow \\ \uparrow & ? & 0 \end{pmatrix} & x\text{-polarized pump} \\ \begin{pmatrix} ? & \uparrow & 0 \\ ? & 0 & \uparrow \end{pmatrix} & y\text{-polarized pump} \end{cases} \tag{19}$$

where zeros (0) indicates entries that must vanish, question marks (?) indicate entries that are irrelevant for the case, and arrows ($\uparrow$) indicate entries that should have large and preferably equal absolute values. These symbolic matrices can be understood as stencils that one must apply to the $\underline{\boldsymbol{\chi}}^{(2)}$ of a given crystal to check if the crystal and its orientation is suitable for the generation of $|\Psi^\pm\rangle$ Bell states.

Since $\underline{\boldsymbol{\chi}}^{(2)}$ is given in the waveguide reference system, one can not only choose a certain crystal class, but also the orientation of the crystal. Crystals of the triclinic, monoclinic, and orthorhombic crystal systems are biaxial, those of the tetragonal, trigonal, and hexagonal crystal systems are uniaxial, and those of the cubic crystal system are isotropic. In total, there are 18 birefringent and 2 isotropic, non-centrosymmetric crystal classes. In the following, we restrict the discussion to the birefringent crystal classes, as the isotropic crystal classes cannot utilize the phase matching scheme discussed here. Nevertheless, we present a brief adapted analysis of the isotropic crystal classes in the Supporting Information.

In uniaxial crystals, the optical axis corresponds to the $\tilde{z}$-axis, and birefringence is best utilized when the $\tilde{z}$-axis is perpendicular to the $z$-axis. Hence, it is advisable to rotate the crystal using the following rotation matrix

$$\mathbf{R}(\gamma, \tilde{\gamma}) = \begin{pmatrix} \sin(\tilde{\gamma})\cos(\gamma) & \sin(\tilde{\gamma})\sin(\gamma) & \cos(\tilde{\gamma}) \\ \cos(\tilde{\gamma})\cos(\gamma) & \cos(\tilde{\gamma})\sin(\gamma) & -\sin(\tilde{\gamma}) \\ -\sin(\gamma) & \cos(\gamma) & 0 \end{pmatrix} \tag{20}$$

where $\gamma$ rotates the crystal around the $z$-axis and $\tilde{\gamma}$ around its own $\tilde{z}$-axis (see Fig. 2n). The susceptibilities in the waveguide reference system ($\chi$) can thus be represented as linear combinations of the susceptibilities in the crystal reference system ($\tilde{\chi}$):

$$\chi_{\ell m}^{(1)} = \sum_{ij} \tilde{\chi}_{ij}^{(1)} R_{i\ell} R_{jm} \tag{21}$$

$$\chi_{\ell m n}^{(2)} = \sum_{ijk} \tilde{\chi}_{ijk}^{(2)} R_{i\ell} R_{jm} R_{kn} \tag{22}$$

Rotations in $\frac{\pi}{2}$-steps are of particular interest here, as they correspond to a permutation of the coordinate axes, and the crystal layers are usually easy to fabricate in these orientations. Table 1 shows $\underline{\boldsymbol{\chi}}^{(2)}$ for such rotations and reveals that an additional rotation by $\pm\frac{\pi}{2}$ along $\gamma$ possesses the following symmetry:

$$\begin{pmatrix} 0 & ? & \uparrow \\ \uparrow & ? & 0 \end{pmatrix} \xrightarrow{\gamma \pm \frac{\pi}{2}} \mp \begin{pmatrix} ? & \uparrow & 0 \\ ? & 0 & \uparrow \end{pmatrix} \tag{23}$$

**Table 1 |** The reduced second-order susceptibility in the waveguide reference system $\underline{\boldsymbol{\chi}}^{(2)}$ for rotations $\mathbf{R}(\gamma,\tilde{\gamma})$ of birefringent crystals in $\frac{\pi}{2}$-steps, where the stacked symbols in front/inside the matrices correspond to those in the row/column headers. Entries of the form $\pm ijk$ are shorthand notations for $\pm\tilde{\chi}^{(2)}_{ijk}$, the second-order susceptibility in the crystal reference system. Note that the intrinsic permutation symmetry $ijk = ikj$ is assumed, while the last row shows which entries are identical when the Kleinman symmetry applies.

| $\underline{\boldsymbol{\chi}}^{(2)}$ | $\tilde{\gamma} = \{\begin{smallmatrix}0\\\pi\end{smallmatrix}\}$ | $\tilde{\gamma} = \pm\frac{\pi}{2}$ |
|---|---|---|
| $\gamma = \{\begin{smallmatrix}0\\\pi\end{smallmatrix}\}$ | $\pm\begin{pmatrix} \pm yyy & \pm yzz & yyz \\ zyy & zzz & \pm zyz \end{pmatrix}$ | $\pm\begin{pmatrix} \pm xxx & \pm xzz & xxz \\ zxx & zzz & \pm zxz \end{pmatrix}$ |
| $\gamma = \pm\frac{\pi}{2}$ | $\mp\begin{pmatrix} zzz & zyy & \mp zyz \\ \mp yzz & \mp yyy & yyz \end{pmatrix}$ | $\mp\begin{pmatrix} zzz & zxx & \mp zxz \\ \mp xzz & \mp xxx & xxz \end{pmatrix}$ |
| Kleinm. symm. | $yyz = zyy$, $yzz = zyz$ | $xzz = zxz$, $xxz = zxx$ |

In words, this means that if a crystal is suitable for one pump polarization, it is automatically suitable for the other as well. The crystal only needs to be rotated by $\gamma \pm \frac{\pi}{2}$. Consequently, it suffices to consider only $\gamma = 0$ in the following discussion. It is also noteworthy that it is beneficial if the Kleinman symmetry applies, since this naturally provides equal values for the ↑ entries of the stencils. The $\tilde{\boldsymbol{\chi}}^{(2)}$ and $\underline{\boldsymbol{\chi}}^{(2)}$ for all birefringent, non-centrosymmetric crystal classes are listed in Table S2 (Supporting Information). Here, we summarize by noting that the application of the stencils to $\underline{\boldsymbol{\chi}}^{(2)}$ results in the following categorization of crystal classes regarding their principal suitability for the generation of $|\Psi^\pm\rangle$ Bell states:

Suitable: 2, m, mm2, 4, $\bar{4}$, 4mm, 3m, 6, 6mm
Unsuitable: 1, 3, 222, 422, $\bar{4}$2m, 32, $\bar{6}$, $\bar{6}$m2, 622

Note that in both categories, the actual suitability of a given crystal ultimately depends on its specific $\tilde{\boldsymbol{\chi}}^{(2)}$ values and the considered crystal rotations. A non-exhaustive list of suitable crystals with large $\underline{\boldsymbol{\chi}}^{(2)}$ components is provided in Table S4 (Supporting Information). In the selection of a crystal, one should further consider technological and socioeconomic factors, such as the availability and CMOS-compatibility of the involved chemical elements and fabrication processes, as well as their potential health hazards.

## 5 SHG Simulations

Two promising material platforms for integrated photonic circuits are lithium niobate[28,29] (LiNbO$_3$, crystal class: 3m) and barium titanate[30,31] (BaTiO$_3$, crystal class: 4mm), who's $\underline{\boldsymbol{\chi}}^{(2)}$ are shown in Table 2. By comparison with the stencil from Eq. (19) it is observed that LiNbO$_3$ has the largest absolute susceptibility component, but in irrelevant positions (?), while BaTiO$_3$ offers significantly larger components in the relevant positions (↑), making it the preferred material.

To corroborate this analysis, we simulated SHG in waveguides made from either BaTiO$_3$ or LiNbO$_3$ with different core dimensions. We also investigated two different crystal orientations which exploit the birefringence of the materials to increase the chance of finding phase matched eigenmodes. The first orientation is given by $\gamma = \tilde{\gamma} = -\frac{\pi}{2}$ (see Eq. (20) and Fig. 2n), which aligns the crystal's $\tilde{\mathbf{z}}$-axis with the waveguide's $\mathbf{x}$-axis. In the following, we will refer to this orientation as $\tilde{\mathbf{z}} \parallel \mathbf{x}$. The second orientation is given by $\gamma = \tilde{\gamma} = 0$, which we will analogously refer to as $\tilde{\mathbf{z}} \parallel \mathbf{y}$.

In Fig. 2, we show that by changing the excitation polarization of the SHG simulations, we can simulate the time-reversed version of the degenerate SPDC process of each of the four basis states in Eq. (3).

The mixed polarization excitation (Figs. 2c, 2f, 2i, 2l) represents the conversion into the $|xy\rangle$ and $|yx\rangle$ basis states; the ones used in the $|\Psi^\pm\rangle$ Bell states. For both materials, there are additional branches of high efficiency which are not found in either of the two corresponding single polarization excitation simulations (Figs. 2a, 2b, 2d, 2e, 2g, 2h, 2j, 2k). This indicates that an efficient nonlinear process in these waveguide geometries requires that the signal and idler photons have orthogonal polarization. In the BaTiO$_3$ simulations (Figs. 2a–2f), these extra branches in the mixed polarization excitation are more efficient than the equivalent branches in the LiNbO$_3$ simulations (Figs. 2g–2l). This is because BaTiO$_3$ has stronger $\chi^{(2)}_{xyy}, \chi^{(2)}_{yxy}$ ($\tilde{\mathbf{z}} \parallel \mathbf{x}$) or $\chi^{(2)}_{xxy}, \chi^{(2)}_{yxx}$ ($\tilde{\mathbf{z}} \parallel \mathbf{y}$) than LiNbO$_3$, allowing for a more efficient conversion process. In principle, if the system were to be operated in reverse from one of these geometries, i.e. the pump source is the phase-matched second harmonic mode from these branches, SPDC could generate the $|\Psi^+\rangle$ Bell state. However, to confirm true Bell state generation, the coefficients $a_{ij}$ need to be found, so that the output state $|\psi\rangle$ can be properly expressed.

**Table 2 |** The reduced second-order susceptibility in the waveguide reference system $\underline{\boldsymbol{\chi}}^{(2)}$ of BaTiO$_3$ and LiNbO$_3$ for the two crystal orientations used in the simulations. Dots represent vanishing entries. The presented values were taken from Weber[32] and are valid for SHG pump wavelengths of $\lambda \approx 1.06$ μm.

| | $\underline{\boldsymbol{\chi}}^{(2)}$ (pm/V) | |
|---|---|---|
| | $\gamma = \tilde{\gamma} = -\frac{\pi}{2}$ ($\tilde{\mathbf{z}} \parallel \mathbf{x}$) | $\gamma = \tilde{\gamma} = 0$ ($\tilde{\mathbf{z}} \parallel \mathbf{y}$) |
| BaTiO$_3$ | $\begin{pmatrix} 13.6 & 31.4 & \cdot \\ \cdot & \cdot & 34.0 \end{pmatrix}$ | $\begin{pmatrix} \cdot & \cdot & 34.0 \\ 31.4 & 13.6 & \cdot \end{pmatrix}$ |
| LiNbO$_3$ | $\begin{pmatrix} -68.0 & -9.8 & \cdot \\ \cdot & \cdot & -9.8 \end{pmatrix}$ | $\begin{pmatrix} 5.2 & \cdot & -9.8 \\ -9.8 & -68.0 & \cdot \end{pmatrix}$ |

## 6 SFG Simulations and Concurrence

The concurrence $C$ quantifies the entanglement of arbitrary two-qubit states[33]

$$C = \left|\langle\psi|\sigma_y \otimes \sigma_y|\overline{\psi}\rangle\right| \quad (24)$$

where $|\psi\rangle$ is a two-qubit state, $\sigma_y = \begin{pmatrix} 0 & -i \\ i & 0 \end{pmatrix}$ one of the Pauli matrices, $\otimes$ the outer product, and $|\overline{\psi}\rangle$ the complex conjugate of $|\psi\rangle$. This expression for the concurrence simplifies for the pure state from Eq. (3):

$$C = 2\left|a_{xx}a_{yy} - a_{xy}a_{yx}\right| \quad (25)$$

The concurrence takes values between 0 and 1, where 1 corresponds to complete entanglement.

The difficulty lies in the determination of the coefficients of the final state $|\psi\rangle$. To avoid the difficulties of a complete quantum-mechanical description, we chose to estimate $C$ via classical wave optics simulations. The key idea is to use the quantum-classical correspondence principle[34–36], where the probabilities of annihilating a signal-idler photon pair via SFG is the same as generating said pair via SPDC.

SFG simulations allow us to determine the concurrence as a function of the signal (and idler) frequency. If we perform four separate SFG simulations, in each of which we excite a pair of modes $|ij\rangle$ (where $i, j \in \{x, y\}$) with a power $P_0 = P(0, \omega_s) + P(0, \omega_i)$ and measure the power $P_{ij} = P(\ell, \omega_p)$ of the generated pump mode at the waveguide end $z = \ell$ (where $P_{ij}$ keeps record of the excitation), we can expect in turn for the SPDC process that $p_{ij} = P_{ij}/P_0$ is the probability for finding the signal-idler pair $|ij\rangle$ associated with the coefficient $a_{ij}$ at the end of the waveguide. Hence, the coefficients of the final state $|\psi\rangle$ can be estimated as

$$|a_{ij}| \approx \frac{\sqrt{p_{ij}}}{\sqrt{p_{xx} + p_{xy} + p_{yx} + p_{yy}}} \quad (26)$$

where we note that this follows the same normalization requirement from Eq. (4).

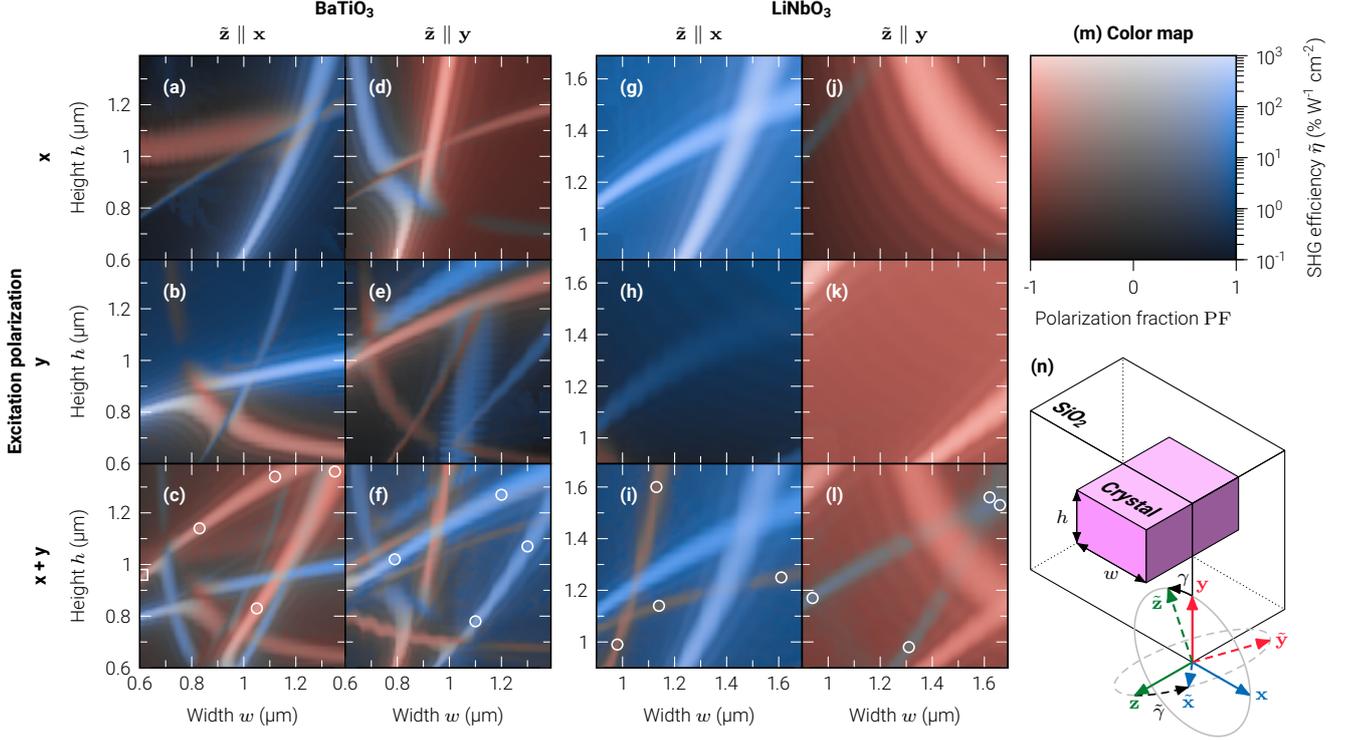

**Fig. 2 | (a–l)** SHG efficiency and polarization fraction, encoded via the **(m)** two-dimensional color map, for (a–f) BaTiO$_3$ and (g–l) LiNbO$_3$ waveguides. **(n)** The crystal reference system ($\tilde{x}, \tilde{y}, \tilde{z}$) is rotated with respect to the waveguide reference system ($x, y, z$), where $\gamma$ rotates the crystal around the $z$-axis and $\tilde{\gamma}$ around its own $\tilde{z}$-axis. The crystalline waveguide with width $w$ and height $h$ is fully embedded in SiO$_2$. In (a–c, g–i), the crystal is rotated by $\gamma = \tilde{\gamma} = -\frac{\pi}{2}$, which means $\tilde{z} \parallel \mathbf{x}$. In (d–f, j–l), the crystal is rotated by $\gamma = \tilde{\gamma} = 0$, which means $\tilde{z} \parallel \mathbf{y}$. The waveguides are excited by (a, d, g, j) the $x$-polarized, (b, e, h, k) the $y$-polarized, or (c, f, i, l) both $x$- and $y$-polarized fundamental eigenmodes. The widths and heights marked by (c, f, i, l) circles and (c) a square relate to Fig. 3 and Fig. 1, respectively.

Once the four coefficients are known, the concurrence can easily be calculated using Eq. (25). We see that for the $|\Psi^\pm\rangle$ Bell states, which have $a_{xy} = \pm a_{yx} = \frac{1}{\sqrt{2}}$ and $a_{xx} = a_{yy} = 0$, the concurrence is equal to 1, showing that these are states with maximum entanglement.

In Fig. 3, we show the results of the SFG sweeps of different geometries previously identified to show promise for $|\Psi^\pm\rangle$ Bell state generation in both BaTiO$_3$ and LiNbO$_3$. For each waveguide, the pump photon wavelength was fixed at $\lambda = 775$ nm, while the signal and idler photon wavelengths were varied. After a waveguide length of $\ell = 100$ μm, the four coefficients for the output state were found using Eq. (26), and the corresponding concurrence was calculated with Eq. (25).

For each geometry, the concurrence at the degenerate frequency is 1 and this then begins to decrease as the signal photon frequency is offset from the degenerate frequency. This makes sense as at the degenerate frequency the signal and idler photons are indistinguishable, and therefore $|a_{xy}|$ and $|a_{yx}|$ must always be equal. As the signal photon frequency increases or decreases, the effective refractive index of the signal and idler photons may change at different rates. This causes the phase matching to get worse between the pump, signal, and idler modes, resulting in the coefficient for that basis state to decrease. This creates a certain bandwidth around the degenerate frequency which supports highly entangled photon pair generation. The exact bandwidth is dependent on the geometry, material, and dimensions of the waveguide, however it is important to note that these simulations show that each of these geometries are capable of being used as integrated Bell state sources near the degenerate frequency they were designed for.

## 7 Conclusion

This study demonstrates that it is possible to efficiently generate polarization-entangled Bell states via SPDC in structurally simple waveguides by leveraging intrinsic crystal properties. Compared to other phase-matching techniques, our approach may require fewer and less complicated fabrication steps, thereby offering a higher integration density and robustness to fabrication tolerances.

To this end, we first identified the components of $\chi^{(2)}$ that enable the generation of cross-polarized photon pairs in monolithic waveguides. We then assumed crystal rotations that maximize the effect of birefringence in uniaxial crystals and are generally easy to fabricate. With this, we categorized all relevant crystal classes regarding their principal suitability and found barium titanate to be superior to lithium niobate, as it offers a higher nonlinear efficiency and a high concurrence over a broader bandwidth.

In summary, our findings highlight the potential for highly efficient and scalable on-chip sources of polarization-entangled photons, paving the way for technological advancements in quantum key distribution and quantum computing technologies. This work not only provides a practical framework for future developments in integrated photonic circuits for quantum applications, but also contributes to the understanding of nonlinear optical processes in waveguides.

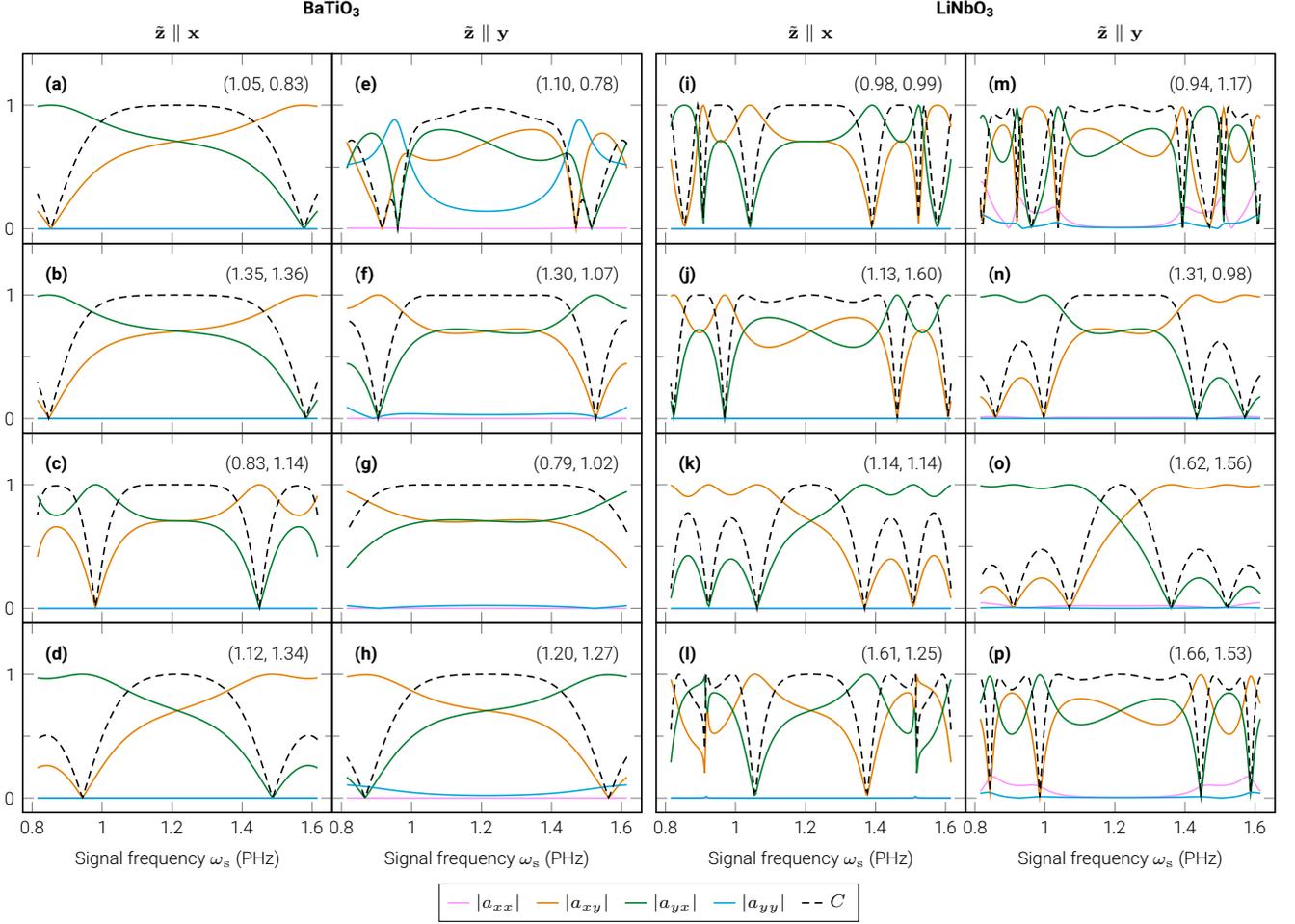

**Fig. 3 | (a–p)** Normalized coefficients $|a_{xx}|$, $|a_{xy}|$, $|a_{yx}|$, $|a_{yy}|$ of the four basis states in the two-qubit state $|\psi\rangle$ as a function of the signal photon frequency $\omega_s$, calculated from the efficiency of the conversion into the pump mode after the corresponding SFG simulation, and the concurrence $C$ of the output state. The waveguide cores were made of (a–h) BaTiO$_3$ or (i–p) LiNbO$_3$, with the crystal rotated such that (a–d, i–l) $\tilde{\mathbf{z}} \parallel \mathbf{x}$ or (e–h, m–p) $\tilde{\mathbf{z}} \parallel \mathbf{y}$. Each waveguide was simulated over a length of $\ell = 100$ µm. The tuple in each panel shows the waveguide width and height $(w, h)$ in micrometers, and corresponds to a circular mark in Fig. 2.

## Methods

### Simulation Methods

From Maxwell's curl equations the vector wave equation can be derived, the solutions of which are a series of eigenvectors with corresponding eigenvalues. Each eigenvector describes how an electric and magnetic field can propagate in the waveguide, and the corresponding eigenvalue is known as the propagation constant. The combined electric and magnetic field is known as an eigenmode, and once normalized, eigenmodes of a specific frequency have the unique property that they are all orthonormal to each other. Therefore, any electromagnetic field propagating through a waveguide can be decomposed into a linear superposition of eigenmodes as

$$\mathbf{E}(x,y,z,\omega) = \sum_m A_m(z)\, \mathbf{E}_m(x,y,\omega)\, e^{i\beta_m z} \quad (27)$$

$$\mathbf{H}(x,y,z,\omega) = \sum_m A_m(z)\, \mathbf{H}_m(x,y,\omega)\, e^{i\beta_m z} \quad (28)$$

where $A_m$ is the amplitude of the eigenmode, $\mathbf{E}_m$ and $\mathbf{H}_m$ are the electric and magnetic mode profiles in the transverse plane, $\beta_m$ is the propagation constant of the mode, and $z$ is the current position in the propagation direction.

Using Eq. (27) as an ansatz in the vector wave equation, we find a set of coupled differential equations that describe how each mode amplitude evolves as the mode propagates along the waveguide[37]:

$$\frac{\partial}{\partial z} A_m(z) = i\frac{\omega}{4}\epsilon_0\, e^{-i\beta_m z}$$
$$\int_{\mathbb{R}^2} \overline{\mathbf{E}_m(x,y,\omega)} \cdot \mathbf{P}_{\mathrm{NL}}(x,y,z,\omega)\, \mathrm{d}x\,\mathrm{d}y \quad (29)$$

We then find the eigenmodes of a specific waveguide geometry and numerically solve the corresponding set of coupled differential equations from Eq. (29) as an initial value problem. This results in a list of complex mode amplitudes, each evaluated at each $z$-step of the initial value solver. With these amplitudes, the total $\mathbf{E}$ and $\mathbf{H}$ fields at each frequency can be found in the waveguide using Eqs. (27) and (28).

For the electric field from Eq. (14), the total guided power $P(z)$ at a point $z$ in the waveguide can be decomposed into the contributions $P(z, \omega_q)$ of the modes q $\in$ {p, s, i} as

$$P(z) = P(z, \omega_p) + P(z, \omega_i) + P(z, \omega_s) \quad (30)$$

$$P(z, \omega_q) = \frac{1}{2}\int_{\mathbb{R}^2} \mathrm{Re}\bigl(\mathbf{E}^q(\mathbf{r}, \omega_q) \times \overline{\mathbf{H}^q(\mathbf{r}, \omega_q)}\bigr) \cdot \hat{\mathbf{z}}\, \mathrm{d}x\,\mathrm{d}y \quad (31)$$

Since SPDC is a purely quantum process, it cannot be simulated directly. However, we can instead simulate the time-reversed processes, namely SHG and SFG, and from those results infer the results of the SPDC process. In these time-reversed processes, the signal and idler photon become the input to the simulation, while the pump photon becomes the output. Without loss of generality, it can be assumed that the waveguide starts at $z = 0$ and is excited there with a certain signal power $P(0, \omega_\text{s})$ and idler power $P(0, \omega_\text{i})$. The normalized efficiency $\eta$ with which pump photons are generated is then defined as

$$\eta(z) := \frac{P(z, \omega_\text{p})}{P(0, \omega_\text{i})\, P(0, \omega_\text{s})} \quad (32)$$

with units of $[\eta] = \text{W}^{-1}$. Since $P(z, \omega_\text{p})$ is generally proportional to $P(0, \omega_\text{i})\, P(0, \omega_\text{s})\, z^2$, i.e. dependent on the square of the waveguide length, we can further define a normalized efficiency $\tilde{\eta}$

$$\tilde{\eta} := \arg\min_a \int_0^\ell \left[\eta(z) - a z^2\right]^2 \text{d}z = \frac{5}{\ell^5} \int_0^\ell z^2 \eta(z)\, \text{d}z \quad (33)$$

with units of $[\eta] = \text{W}^{-1}\text{m}^{-2}$, where $\tilde{\eta} z^2$ fits the normalized efficiency $\eta(z)$ in the least squares sense and $\ell$ is the actual length of the waveguide. Values of equivalent quantities are often given in literature in units of $\%\,\text{W}^{-1}\text{cm}^{-2}$.

Moreover, we define the polarization fraction PF as

$$\text{PF}(\omega) = \frac{\int_{\mathbb{R}^2} |E_x(x,y,\omega)|^2 - |E_y(x,y,\omega)|^2\, \text{d}x\, \text{d}y}{\int_{\mathbb{R}^2} |\mathbf{E}(x,y,\omega)|^2\, \text{d}x\, \text{d}y} \quad (34)$$

which indicates whether the guided electric field is predominantly $x$-polarized ($\text{PF} = 1$) or predominantly $y$-polarized ($\text{PF} = -1$).

## Acknowledgments

The authors acknowledge the Fraunhofer Attract Grant SILIQUA No. 40-04866, the BMBF projects MEXSIQUO Grant No. 13N16967 and SINNER Grant No. 16KIS1792, and the Collaborative Research Center (CRC/SFB) 1375 NOA.

## Conflict of Interest

The authors declare no conflict of interest.

## Data Availability Statement

The data that support the findings of this study are available from the corresponding author upon reasonable request.

## S1 Contracted Kleinman notation

To enhance the readability in the following tables, we employ the contracted Kleinman notation[1] $\chi^{(2)}_{ijk} = \chi^{(2)}_{i\ell}$, where the indices $x, y, z$ are enumerated by $1, 2, 3$ and the last pair of indices $jk$ maps to $\ell$ according to:

$$\begin{array}{c|cccccc} jk & 11 & 22 & 33 & 23,32 & 13,31 & 12,21 \\ \ell & 1 & 2 & 3 & 4 & 5 & 6 \end{array} \quad \text{(S1)}$$

## S2 Second-order susceptibility of birefringent and non-centrosymmetric crystals

Table S1 indicates how the entries of $\underline{\chi}^{(2)}$ permute when the crystal is rotated by $\mathbf{R}(\gamma, \tilde{\gamma})$ in steps of $\frac{\pi}{2}$ (cf. Fig. S1a). Table S2 shows the general structure of both $\tilde{\underline{\chi}}^{(2)}$ and $\underline{\chi}^{(2)}$ for all birefringent and non-centrosymmetric crystal classes. Table S4 lists the values of $\underline{\chi}^{(2)}$ for selected crystals and rotations $\mathbf{R}(\gamma = 0, \tilde{\gamma})$. Note that the values for other rotations can be deduced from Table S1.

## S3 Second-order susceptibility of isotropic and non-centrosymmetric crystals

The cubic crystal class, and some of the birefringent crystal classes (e.g. 222 and $\bar{4}$2m), can still be used for the generation of $|\Psi^\pm\rangle$ Bell states if the crystal is rotated in another manner. For example, the following rotation matrix $\mathbf{R}(\phi)$ is particularly well suited for the cubic crystal class

$$\mathbf{R}(\phi) = \mathbf{P} \cdot \begin{pmatrix} \frac{1}{\sqrt{2}}\cos(\phi) & \frac{1}{\sqrt{2}}\sin(\phi) & \frac{1}{\sqrt{2}} \\ -\frac{1}{\sqrt{2}}\cos(\phi) & -\frac{1}{\sqrt{2}}\sin(\phi) & \frac{1}{\sqrt{2}} \\ \sin(\phi) & -\cos(\phi) & 0 \end{pmatrix} \quad \text{(S2)}$$

where $\mathbf{P}$ is an arbitrary permutation matrix that sets one of the $(110), (011), (101)$ crystal planes perpendicular to the $z$-axis, and $\phi$ rotates the selected crystal plane round the $z$-axis (cf. Fig. S1b).

Table S3 shows that the two applicable cubic crystal classes ($23, \bar{4}$3m) are, when rotated by $\mathbf{R}(\phi)$, well suited for the generation of $|\Psi^\pm\rangle$ Bell states. Lastly, Table S5 shows selected isotropic crystals with high values for $\tilde{\chi}^{(2)}_{14}$.

## References

[1] R. W. Boyd, "Nonlinear optics", Elsevier, **2020**.

[2] M. J. Weber, "Handbook of Optical Materials", CRC Press, **2018**.

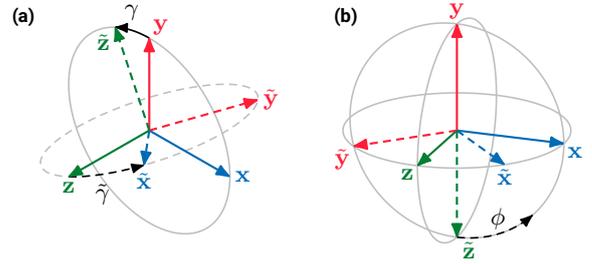

**Fig. S1** | The coordinate systems and rotations used for **(a)** birefringent and **(b)** isotropic crystal classes. The crystal reference system $(\tilde{x}, \tilde{y}, \tilde{z})$ lies rotated in the waveguide reference system $(x, y, z)$. In (a), $\gamma$ rotates the crystal around the $z$-axis and $\tilde{\gamma}$ around its own $\tilde{z}$-axis. In (b), $\phi$ rotates the crystal around the $z$-axis and, in the illustrated case, the permutation matrix $\mathbf{P}$ in $\mathbf{R}(\phi)$ was set to the identity matrix $\mathbf{I}$.

**Table S1** | The reduced second-order susceptibility in the waveguide reference system $\underline{\chi}^{(2)}$ for rotations $\mathbf{R}(\gamma, \tilde{\gamma})$ of birefringent crystals in $\frac{\pi}{2}$-steps, where the stacked symbols in front/inside the matrices correspond to those in the row/column headers. Entries of the form $\pm i\ell$ are shorthand notations for $\pm \tilde{\chi}^{(2)}_{i\ell}$, the second-order susceptibility in the crystal reference system. The last row shows which entries are identical when the Kleinman symmetry applies.

| $\underline{\chi}^{(2)}(\gamma, \tilde{\gamma})$ | $\tilde{\gamma} = \{^0_\pi\}$ | $\tilde{\gamma} = \pm\frac{\pi}{2}$ |
|---|---|---|
| $\gamma = \{^0_\pi\}$ | $\pm \begin{pmatrix} \pm 22 & \pm 23 & 24 \\ 32 & 33 & \pm 34 \\ 33 & 32 & \mp 34 \\ \mp 23 & \mp 22 & 24 \end{pmatrix}$ | $\pm \begin{pmatrix} \pm 11 & \pm 13 & 15 \\ 31 & 33 & \pm 35 \\ 33 & 31 & \mp 35 \\ \mp 13 & \mp 11 & 15 \end{pmatrix}$ |
| $\gamma = \pm\frac{\pi}{2}$ | $\mp$ | $\mp$ |
| Kleinman symmetry | 24 = 32, 23 = 34 | 13 = 35, 15 = 31 |

**Table S2** | The second-order susceptibility in the crystal reference system $\tilde{\boldsymbol{\chi}}^{(2)}$, and in reduced form in the waveguide reference system $\underline{\boldsymbol{\chi}}^{(2)}$, for all birefringent and non-centrosymmetric crystal classes. Entries of the form $\pm i\ell$ are shorthand notations for $\pm\tilde{\chi}^{(2)}_{i\ell}$. Dots represent vanishing entries. The reduced form $\underline{\boldsymbol{\chi}}^{(2)}$ is shown for selected crystal rotations $\mathbf{R}(\gamma=0,\tilde{\gamma})$, where the stacked symbols inside the matrices correspond to those in the column headers. The reduced form for other values of $\gamma$ can be derived from Table S1 via symmetry considerations. The last row shows which entries are identical when the Kleinman symmetry applies.

| Crystal system | Crystal class | | $\tilde{\boldsymbol{\chi}}^{(2)}$ $\pm i\ell \to \pm\tilde{\chi}^{(2)}_{i\ell}$ | $\underline{\boldsymbol{\chi}}^{(2)}(\gamma=0,\tilde{\gamma})$ $\tilde{\gamma}=\{{}^0_\pi$ | $\tilde{\gamma}=\pm\frac{\pi}{2}$ |
|---|---|---|---|---|---|
| Triclinic (biaxial) | 1 | $C_1$ | $\begin{pmatrix} 11 & 12 & 13 & 14 & 15 & 16 \\ 21 & 22 & 23 & 24 & 25 & 26 \\ 31 & 32 & 33 & 34 & 35 & 36 \end{pmatrix}$ | $\begin{pmatrix} \pm 22 & \pm 23 & 24 \\ 32 & 33 & \pm 34 \end{pmatrix}$ | $\begin{pmatrix} \pm 11 & \pm 13 & 15 \\ 31 & 33 & \pm 35 \end{pmatrix}$ |
| Monoclinic (biaxial) | 2 | $C_2$ | $\begin{pmatrix} \cdot & \cdot & \cdot & 14 & \cdot & 16 \\ 21 & 22 & 23 & \cdot & 25 & \cdot \\ \cdot & \cdot & \cdot & 34 & \cdot & 36 \end{pmatrix}$ | $\begin{pmatrix} \pm 22 & \pm 23 & \cdot \\ \cdot & \cdot & \pm 34 \end{pmatrix}$ | $\begin{pmatrix} \cdot & \cdot & \cdot \\ \cdot & \cdot & \cdot \end{pmatrix}$ |
| | m | $C_{1h}$ | $\begin{pmatrix} 11 & 12 & 13 & \cdot & 15 & \cdot \\ \cdot & \cdot & \cdot & 24 & \cdot & 26 \\ 31 & 32 & 33 & \cdot & 35 & \cdot \end{pmatrix}$ | $\begin{pmatrix} \cdot & \cdot & 24 \\ 32 & 33 & \cdot \end{pmatrix}$ | $\begin{pmatrix} \pm 11 & \pm 13 & 15 \\ 31 & 33 & \pm 35 \end{pmatrix}$ |
| Ortho-rhombic (biaxial) | 222 | $D_2$ | $\begin{pmatrix} \cdot & \cdot & \cdot & 14 & \cdot & \cdot \\ \cdot & \cdot & \cdot & \cdot & 25 & \cdot \\ \cdot & \cdot & \cdot & \cdot & \cdot & 36 \end{pmatrix}$ | $\begin{pmatrix} \cdot & \cdot & \cdot \\ \cdot & \cdot & \cdot \end{pmatrix}$ | $\begin{pmatrix} \cdot & \cdot & \cdot \\ \cdot & \cdot & \cdot \end{pmatrix}$ |
| | mm2 | $C_{2v}$ | $\begin{pmatrix} \cdot & \cdot & \cdot & \cdot & 15 & \cdot \\ \cdot & \cdot & \cdot & 24 & \cdot & \cdot \\ 31 & 32 & 33 & \cdot & \cdot & \cdot \end{pmatrix}$ | $\begin{pmatrix} \cdot & \cdot & 24 \\ 32 & 33 & \cdot \end{pmatrix}$ | $\begin{pmatrix} \cdot & \cdot & 15 \\ 31 & 33 & \cdot \end{pmatrix}$ |
| Tetragonal (uniaxial) | 4 | $C_4$ | $\begin{pmatrix} \cdot & \cdot & \cdot & 14 & 15 & \cdot \\ \cdot & \cdot & \cdot & 15 & -14 & \cdot \\ 31 & 31 & 33 & \cdot & \cdot & \cdot \end{pmatrix}$ | $\begin{pmatrix} \cdot & \cdot & 15 \\ 31 & 33 & \cdot \end{pmatrix}$ | $\begin{pmatrix} \cdot & \cdot & 15 \\ 31 & 33 & \cdot \end{pmatrix}$ |
| | $\bar{4}$ | $S_4$ | $\begin{pmatrix} \cdot & \cdot & \cdot & 14 & 15 & \cdot \\ \cdot & \cdot & \cdot & -15 & 14 & \cdot \\ 31 & -31 & \cdot & \cdot & \cdot & 36 \end{pmatrix}$ | $\begin{pmatrix} \cdot & \cdot & -15 \\ -31 & \cdot & \cdot \end{pmatrix}$ | $\begin{pmatrix} \cdot & \cdot & 15 \\ 31 & \cdot & \cdot \end{pmatrix}$ |
| | 422 | $D_4$ | $\begin{pmatrix} \cdot & \cdot & \cdot & 14 & \cdot & \cdot \\ \cdot & \cdot & \cdot & \cdot & -14 & \cdot \\ \cdot & \cdot & \cdot & \cdot & \cdot & \cdot \end{pmatrix}$ | $\begin{pmatrix} \cdot & \cdot & \cdot \\ \cdot & \cdot & \cdot \end{pmatrix}$ | $\begin{pmatrix} \cdot & \cdot & \cdot \\ \cdot & \cdot & \cdot \end{pmatrix}$ |
| | 4mm | $C_{4v}$ | $\begin{pmatrix} \cdot & \cdot & \cdot & \cdot & 15 & \cdot \\ \cdot & \cdot & \cdot & 15 & \cdot & \cdot \\ 31 & 31 & 33 & \cdot & \cdot & \cdot \end{pmatrix}$ | $\begin{pmatrix} \cdot & \cdot & 15 \\ 31 & 33 & \cdot \end{pmatrix}$ | $\begin{pmatrix} \cdot & \cdot & 15 \\ 31 & 33 & \cdot \end{pmatrix}$ |
| | $\bar{4}2m$ | $D_{2d}$ | $\begin{pmatrix} \cdot & \cdot & \cdot & 14 & \cdot & \cdot \\ \cdot & \cdot & \cdot & \cdot & 14 & \cdot \\ \cdot & \cdot & \cdot & \cdot & \cdot & 36 \end{pmatrix}$ | $\begin{pmatrix} \cdot & \cdot & \cdot \\ \cdot & \cdot & \cdot \end{pmatrix}$ | $\begin{pmatrix} \cdot & \cdot & \cdot \\ \cdot & \cdot & \cdot \end{pmatrix}$ |
| Trigonal (uniaxial) | 3 | $C_3$ | $\begin{pmatrix} 11 & -11 & \cdot & 14 & 15 & -22 \\ -22 & 22 & \cdot & 15 & -14 & -11 \\ 31 & 31 & 33 & \cdot & \cdot & \cdot \end{pmatrix}$ | $\begin{pmatrix} \pm 22 & \cdot & 15 \\ 31 & 33 & \cdot \end{pmatrix}$ | $\begin{pmatrix} \pm 11 & \cdot & 15 \\ 31 & 33 & \cdot \end{pmatrix}$ |
| | 32 | $D_3$ | $\begin{pmatrix} 11 & -11 & \cdot & 14 & \cdot & \cdot \\ \cdot & \cdot & \cdot & \cdot & -14 & -11 \\ \cdot & \cdot & \cdot & \cdot & \cdot & \cdot \end{pmatrix}$ | $\begin{pmatrix} \cdot & \cdot & \cdot \\ \cdot & \cdot & \cdot \end{pmatrix}$ | $\begin{pmatrix} \pm 11 & \cdot & \cdot \\ \cdot & \cdot & \cdot \end{pmatrix}$ |
| | 3m | $C_{3v}$ | $\begin{pmatrix} \cdot & \cdot & \cdot & \cdot & 15 & -22 \\ -22 & 22 & \cdot & 15 & \cdot & \cdot \\ 31 & 31 & 33 & \cdot & \cdot & \cdot \end{pmatrix}$ | $\begin{pmatrix} \pm 22 & \cdot & 15 \\ 31 & 33 & \cdot \end{pmatrix}$ | $\begin{pmatrix} \cdot & \cdot & 15 \\ 31 & 33 & \cdot \end{pmatrix}$ |
| Hexagonal (uniaxial) | $\bar{6}$ | $C_{3h}$ | $\begin{pmatrix} 11 & -11 & \cdot & \cdot & \cdot & -22 \\ -22 & 22 & \cdot & \cdot & \cdot & -11 \\ \cdot & \cdot & \cdot & \cdot & \cdot & \cdot \end{pmatrix}$ | $\begin{pmatrix} \pm 22 & \cdot & \cdot \\ \cdot & \cdot & \cdot \end{pmatrix}$ | $\begin{pmatrix} \pm 11 & \cdot & \cdot \\ \cdot & \cdot & \cdot \end{pmatrix}$ |
| | $\bar{6}m2$ | $D_{3h}$ | $\begin{pmatrix} \cdot & \cdot & \cdot & \cdot & \cdot & -22 \\ -22 & 22 & \cdot & \cdot & \cdot & \cdot \\ \cdot & \cdot & \cdot & \cdot & \cdot & \cdot \end{pmatrix}$ | $\begin{pmatrix} \pm 22 & \cdot & \cdot \\ \cdot & \cdot & \cdot \end{pmatrix}$ | $\begin{pmatrix} \cdot & \cdot & \cdot \\ \cdot & \cdot & \cdot \end{pmatrix}$ |
| | 6 | $C_6$ | see: 4 | | |
| | 622 | $D_6$ | see: 422 | | |
| | 6mm | $C_{6v}$ | see: 4mm | | |
| Kleinman symmetry | | | 12 = 26, 13 = 35, 14 = 25 = 36, 15 = 31, 16 = 21, 23 = 34, 24 = 32 | | |

**Table S3** | The second-order susceptibility in the crystal reference system $\tilde{\boldsymbol{\chi}}^{(2)}$, and in reduced form in the waveguide reference system $\underline{\boldsymbol{\chi}}^{(2)}$, for all isotropic and non-centrosymmetric crystal classes. Entries of the form $\pm i\ell$ are shorthand notations for $\pm\tilde{\chi}^{(2)}_{i\ell}$. Dots represent vanishing entries. The reduced form $\underline{\boldsymbol{\chi}}^{(2)}$ is shown for selected rotations $\mathbf{R}(\phi)$, where the stacked symbols inside the matrices correspond to those in the column headers. Note that the presented rotations are independent of the permutation matrix $\mathbf{P}$.

| Crystal system | Crystal class | | $\tilde{\boldsymbol{\chi}}^{(2)}$ $\pm i\ell \to \pm\tilde{\chi}^{(2)}_{i\ell}$ | $\underline{\boldsymbol{\chi}}^{(2)}(\phi)$ $\phi = \{{0 \atop \pi}$ | $\phi = \pm\frac{\pi}{2}$ |
|---|---|---|---|---|---|
| Cubic | 23 | T | $\begin{pmatrix} \cdot & \cdot & \cdot & 14 & \cdot & \cdot \\ \cdot & \cdot & \cdot & \cdot & 14 & \cdot \\ \cdot & \cdot & \cdot & \cdot & \cdot & 14 \end{pmatrix}$ | $\begin{pmatrix} \cdot & \cdot & \pm 14 \\ \pm 14 & \cdot & \cdot \\ \cdot & \cdot & \cdot \end{pmatrix}$ | $\begin{pmatrix} \cdot & \mp 14 & \cdot \\ \cdot & \cdot & \cdot \\ \cdot & \cdot & \mp 14 \end{pmatrix}$ |
| | $\bar{4}3m$ | $T_d$ | see: 23 | | |

**Table S4** | The reduced second-order susceptibility of selected birefringent crystals in the waveguide reference system $\underline{\chi}^{(2)}$, scaled by a factor of $\frac{1}{2}$, for selected crystal rotations $\mathbf{R}(\gamma = 0, \tilde{\gamma})$. The presented values were taken from Weber[2] and are valid for SHG pump wavelengths of $\lambda \approx 1.06$ µm.

| Crystal class | Chemical formula | $\frac{1}{2}\underline{\chi}^{(2)}(\gamma = 0, \tilde{\gamma})$ (pm/V) | $\tilde{\gamma}$ |
|---|---|---|---|
| mm2 | PbNb$_2$O$_{11}$ | $\begin{pmatrix} \cdot & \cdot & -5.4 \\ -5.9 & -8.9 & \cdot \end{pmatrix}$ | $0$ |
|  |  | $\begin{pmatrix} \cdot & \cdot & 5.9 \\ 6.5 & -8.9 & \cdot \end{pmatrix}$ | $\frac{\pi}{2}$ |
|  | LiInO$_2$ | $\begin{pmatrix} \cdot & \cdot & 8.6 \\ 8.6 & 15.8 & \cdot \end{pmatrix}$ | $0$ |
|  |  | $\begin{pmatrix} \cdot & \cdot & 9.9 \\ 9.9 & 15.8 & \cdot \end{pmatrix}$ | $\frac{\pi}{2}$ |
|  | KTiOPO$_4$ | $\begin{pmatrix} \cdot & \cdot & 7.6 \\ 5.0 & 13.7 & \cdot \end{pmatrix}$ | $0$ |
|  |  | $\begin{pmatrix} \cdot & \cdot & 6.1 \\ 6.5 & 13.7 & \cdot \end{pmatrix}$ | $\frac{\pi}{2}$ |
|  | KNbO$_3$ | $\begin{pmatrix} \cdot & \cdot & 11.3 \\ 11.3 & -19.6 & \cdot \end{pmatrix}$ | $0$ |
|  |  | $\begin{pmatrix} \cdot & \cdot & -12.9 \\ -12.9 & -19.6 & \cdot \end{pmatrix}$ | $\frac{\pi}{2}$ |
|  | Ba$_2$NaNb$_5$O$_{15}$ | $\begin{pmatrix} \cdot & \cdot & -12.8 \\ -12.8 & -17.6 & \cdot \end{pmatrix}$ | $0$ |
|  |  | $\begin{pmatrix} \cdot & \cdot & -12.8 \\ -12.8 & -17.6 & \cdot \end{pmatrix}$ | $\frac{\pi}{2}$ |
| $\bar{4}$ | InPS$_4$ | $\begin{pmatrix} \cdot & \cdot & -26.3 \\ -26.3 & \cdot & \cdot \end{pmatrix}$ | $0$ |
|  |  | $\begin{pmatrix} \cdot & \cdot & 26.3 \\ 26.3 & \cdot & \cdot \end{pmatrix}$ | $\frac{\pi}{2}$ |
| 4mm | SrBaNb$_5$O$_{15}$ | $\begin{pmatrix} \cdot & \cdot & 6.0 \\ 4.3 & 11.3 & \cdot \end{pmatrix}$ | $0$ |
|  | PbTiO$_3$ | $\begin{pmatrix} \cdot & \cdot & 33.3 \\ 37.6 & 7.5 & \cdot \end{pmatrix}$ | $0$ |
|  | BaTiO$_3$ | $\begin{pmatrix} \cdot & \cdot & 17.0 \\ 15.7 & 6.8 & \cdot \end{pmatrix}$ | $0$ |
| 3m | LiTaO$_3$ | $\begin{pmatrix} 1.7 & \cdot & -1.1 \\ -1.1 & -16.4 & \cdot \end{pmatrix}$ | $0$ |
|  |  | $\begin{pmatrix} \cdot & \cdot & -1.1 \\ -1.1 & -16.4 & \cdot \end{pmatrix}$ | $\frac{\pi}{2}$ |
|  | LiNbO$_3$ | $\begin{pmatrix} 2.6 & \cdot & -4.9 \\ -4.9 & -34.0 & \cdot \end{pmatrix}$ | $0$ |
|  |  | $\begin{pmatrix} \cdot & \cdot & -4.9 \\ -4.9 & -34.0 & \cdot \end{pmatrix}$ | $\frac{\pi}{2}$ |
| 6 | LiIO$_3$ | $\begin{pmatrix} \cdot & \cdot & -5.0 \\ -5.0 & -5.2 & \cdot \end{pmatrix}$ | $0$ |
| 6mm | CdS | $\begin{pmatrix} \cdot & \cdot & 28.9 \\ -26.4 & 44.0 & \cdot \end{pmatrix}$ | $0$ |
|  | SiC | $\begin{pmatrix} \cdot & \cdot & 8.0 \\ 8.6 & -14.4 & \cdot \end{pmatrix}$ | $0$ |
|  | $\alpha$-ZnS | $\begin{pmatrix} \cdot & \cdot & 6.7 \\ -7.6 & 13.8 & \cdot \end{pmatrix}$ | $0$ |

**Table S5** | The second-order susceptibility of selected isotropic crystals in the crystal reference system $\tilde{\chi}_{14}^{(2)}$, scaled by a factor of $\frac{1}{2}$. The presented values were taken from Weber[2] and are valid for the indicated SHG pump wavelengths $\lambda$.

| Crystal class | Chemical formula | $\frac{1}{2}\tilde{\chi}_{14}^{(2)}$ (pm/V) | Wavelength $\lambda$ (µm) |
|---|---|---|---|
| $\bar{4}3m$ | CdTe | 167 | 10.60 |
|  | GaAs | 209 | 1.06 |
|  | GaP | 72 | 1.06 |
|  | GaSb | 628 | 10.60 |
|  | InAs | 364 | 1.06 |
|  | InP | 143 | 1.06 |
|  | InSb | 520 | 1.06 |
|  | ZnTe | 90 | 1.06 |